\documentclass[twocolumn]{aastex631}
\usepackage{natbib}
\usepackage{amsmath}
\usepackage{bm}
\usepackage{graphicx}
\usepackage{multirow}
\usepackage{booktabs} 
\usepackage{array}
\usepackage{xcolor} 
\usepackage{hyperref}
\usepackage{soul}

\begin{document}

\title{A Data-constrained Magnetohydrodynamic Simulation of Successive X-class Flares in Solar Active Region 13842. II. Dynamics of the Solar Eruption Associated with the X9.0 Solar Flare}

\author[0000-0003-2002-0247]{Keitarou Matsumoto}
\affiliation{Center for Solar-Terrestrial Research, New Jersey Institute of Technology, University Heights, Newark, NJ 07102-1982, USA}
\author[0000-0001-5121-5122]{Satoshi Inoue}
\affiliation{Center for Solar-Terrestrial Research, New Jersey Institute of Technology, University Heights, Newark, NJ 07102-1982, USA}
\author[0000-0001-9046-6688]{Keiji Hayashi}
\affiliation{Center for Solar-Terrestrial Research, New Jersey Institute of Technology, University Heights, Newark, NJ 07102-1982, USA}
\author[0000-0002-6018-3799]{Nian Liu}
\affiliation{Center for Solar-Terrestrial Research, New Jersey Institute of Technology, University Heights, Newark, NJ 07102-1982, USA}
\author[0009-0008-7623-0140]{Ying Wang}
\affiliation{Center for Solar-Terrestrial Research, New Jersey Institute of Technology, University Heights, Newark, NJ 07102-1982, USA}
\author[0000-0002-5865-7924]{Jeongwoo Lee}
\affiliation{Center for Solar-Terrestrial Research, New Jersey Institute of Technology, University Heights, Newark, NJ 07102-1982, USA}
\author[0000-0002-8179-3625]{Ju Jing}
\affiliation{Center for Solar-Terrestrial Research, New Jersey Institute of Technology, University Heights, Newark, NJ 07102-1982, USA}
\author[0000-0002-5233-565X]{Haimin Wang}
\affiliation{Center for Solar-Terrestrial Research, New Jersey Institute of Technology, University Heights, Newark, NJ 07102-1982, USA}

\email{km876@njit.edu}



\begin{abstract}
Active region NOAA 13842 produced two successive solar flares: an X7.1-class flare on October 1, 2024, and an X9.0-class flare on October 3, 2024. This study continues our previous simulation work that successfully reproduced the X7.1-class solar flare (\citealt{Matsumoto2025}). In this study, we performed a data-constrained magnetohydrodynamic (MHD) simulation using the nonlinear force-free field (NLFFF) as the initial condition to investigate the X9.0-class solar flare. The NLFFF showed the sheared field lines, resulting in the tether-cutting reconnection, the magnetic flux ropes (MFRs), and eventually led to eruption. The magnetic reconnection during the pre-eruption phase plays a critical role in accelerating the subsequent eruption, which is driven by torus instability and magnetic reconnection. Furthermore, our simulation results are consistent with several observational features associated with the X9.0 flare. This simulation could reproduce diverse phenomena associated with the X9.0 flare, including the tether-cutting reconnection, the flare ribbons and the flare loops, the transverse field enhancement, and the remote brightening away from the flare ribbons. However, the initial trigger, magnetic flux emergence, was inferred from observations rather than explicitly modeled, and future comprehensive simulations should incorporate this mechanism directly.
\end{abstract}



\section{Introduction} \label{sec:intro}

Solar flares are among the most energetic and dynamic phenomena in our solar system. They suddenly release vast amounts of magnetic energy stored in the solar corona (\citealt{Fletcher2011}), providing an opportunity to deepen our understanding of fundamental plasma physics (\citealt{Shibata2011}). The impact of solar flares, especially when accompanied by Coronal Mass Ejections(CMEs), is significant for space weather conditions of the Earth as they can disrupt satellite communications and various technological systems on Earth (\citealt{Tsurutani2009, Curto2020}). Despite considerable advancements in solar physics, the precise mechanisms that trigger flares remain one of the most important and elusive topics in the field.

Multiple lines of observational evidence indicate that small-scale changes in the magnetic field configuration can contribute to large-scale solar eruptions. Two well-known processes frequently associated with triggering processes are flux cancellation (\citealt{Martin1985}) and flux emergence (\citealt{Chen2000}). Flux cancellation involves the collision and subsequent annihilation of the photospheric magnetic fluxes, leading to localized reconnection that can cause the flares. On the other hand, flux emergence here refers to the ascent of new magnetic flux from the solar interior into localized areas of an active region, where it can interact with the pre-existing coronal fields. As a result, the pre-existing magnetic fields may lose equilibrium and end up with the eruption (\citealt{Kusano2012, Bamba2013, Torok2024}). Both processes have been regarded as precursors to the solar flares and coronal mass ejections (CME), where a magnetic flux rope (MFR) plays a crucial role as the core structure of the CME (\citealt{Forbes2000}).

Several magnetohydrodynamic (MHD) simulations have successfully demonstrated the formation process of the MFR that leads to an eruption, driven by flux cancellation and small-scale flux emergence (\citealt{Amari2003, Kusano2012}). The formation and early evolution of the MFR, and ultimately its eruption, are driven by the magnetic reconnection and the following ideal-MHD instabilities. As a driving process of the MFR, MHD instabilities are possible candidates, such as the torus instability (TI: \citealt{Bateman1978, Kliem2006}), the kink instability (\citealt{Kruskal1954, Torok2005}), and the double-arc instability (\citealt{Ishiguro2017}) have been proposed to trigger and explain the solar eruptions. Even if there is no MFR present in pre-eruption stage, the tether-cutting reconnection model (\citealt{Moore2001}) and breakout models (\citealt{Antiochos1999}) have been proposed as possible mechanisms for initiating eruptions. Some models suggest that coupling processes between magnetic reconnection and MHD instability play an important role in triggering eruption(\citealt{Aulanier2010,Amari2014,Inoue2018, Xing2024}). Despite this progress, the triggering mechanism and the detailed processes of solar eruptions remain subjects of ongoing debate.

To understand them, we conducted a data-constrained MHD simulation that incorporates the observed photospheric magnetic field into the simulation. Since the solar coronal plasma is in a low $\beta$ state, where $\beta$ is the ratio of the gas pressure to the magnetic pressure and typically ranges from 0.01 to 0.1 (\citealt{Gary2001}), the magnetic field can be well approximated as a force-free field. Therefore, nonlinear force-free field (NLFFF) extrapolation based on observational data has proven to be a useful method for reconstructing the three-dimensional (3D) magnetic field(\citealt{wiegelmann2012}). However, since the NLFFF approach provides a static state, it cannot capture the dynamic evolution of the magnetic field during solar flares. To address this limitation, data-constrained MHD simulations have been applied to investigate the onset and evolution of coronal magnetic fields during solar flares (\citealt{Inoue2014b,Inoue2015,Jiang2016, Muhamad2017,Liu2025, Hayashi2018, Matsumoto2025}).

Two major X-class flares occurred in the solar active region (AR) NOAA 13842, the X7.1 flare on October 1, 2024  (SOL2024-10-01T22:20) and the X9.0 flare two days later (SOL2024-10-03T12:18), both of which were associated with CMEs. In \cite{Matsumoto2025}, our MHD simulations successfully replicated the eruption of the X7.1 flare, which was driven by the interaction between TI and reconnection. In this study, we focus on the X9.0 flare. The vector magnetic field data from both the Helioseismic and Magnetic Imager (HMI; \citealt{Scherrer2012}) onboard the Solar Dynamics Observatory (SDO; \citealt{Pesnell2012}) and the Solar Optical Telescope (SOT; \citealt{Tsuneta2008}) onboard the Hinode satellite (\citealt{Kosugi2007}) are used in this study. Among these, the HMI data are used for the NLFFF extrapolation.

Meanwhile, the extreme ultraviolet images are supplied by the Atmospheric Imaging Assembly (AIA; \citealt{Lemen2012}) onboard the SDO. To investigate the 3D magnetic field evolution leading to the X9.0 flare on October 3, 2024, we performed a data-constrained MHD simulation using the NLFFF extrapolation as the initial condition, following the approach used to study the dynamics of the X7.1 flare \citep{Matsumoto2025}. The structure of this study is as follows: Sections \ref{sec:sec2}, \ref{sec:sec3}, and \ref{sec:sec4} cover observations and methods, results, and discussion, respectively. Finally, Section \ref{sec:sec5} summarizes the conclusions.

\section{Observations and Numerical Methods}
\label{sec:sec2}
\subsection{Observations}
\begin{figure*}[ht!]
\plotone{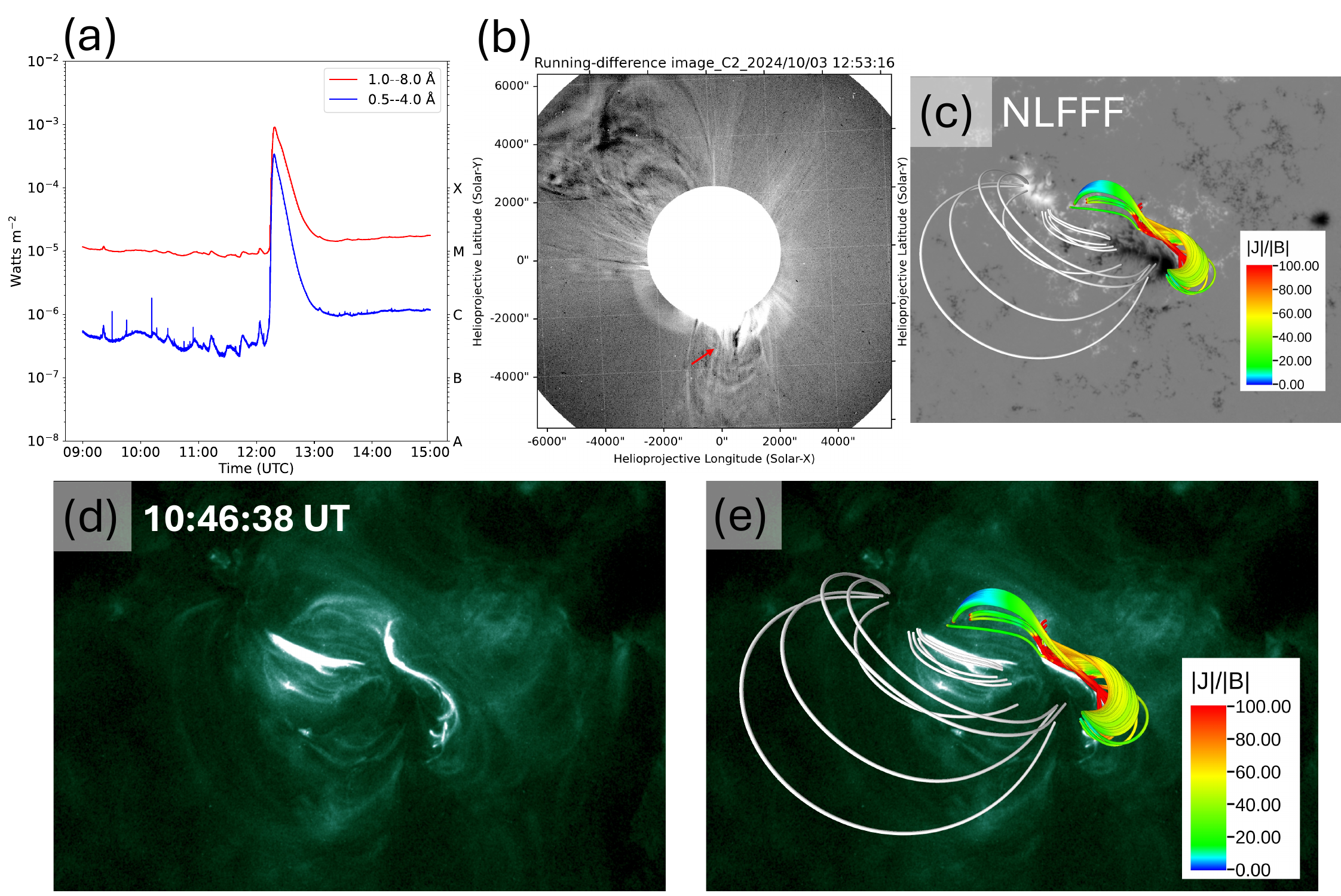}
\caption{(a) Time evolution of the X-ray flux measured by the GOES-16 satellite between 09:00 UT and 15:00 UT on October 3. The solar X-ray emissions in the 1.0–8.0 \AA\ and 0.5–4.0 \AA\ passbands are shown in red and blue, respectively. (b) LASCO/C2 running-difference image observed at 12:53:16 UT. The red arrow shows a CME generated by the X9.0 flare. (c) An NLFFF extrapolated from the vector magnetogram of HMI data at 10:48 UT on 2024 October 3. The background gray scale corresponds to the HMI magnetogram at 10:48 UT. (d) AIA 94 \AA\ image observed at 10:46:38 UT on 2024 October 3. (e) Overlay of the NLFFF extrapolated in (c) with the AIA 94 \AA\ image in (d).
}
\label{fig:fig1}
\end{figure*}
\begin{figure*}[ht!]
\plotone{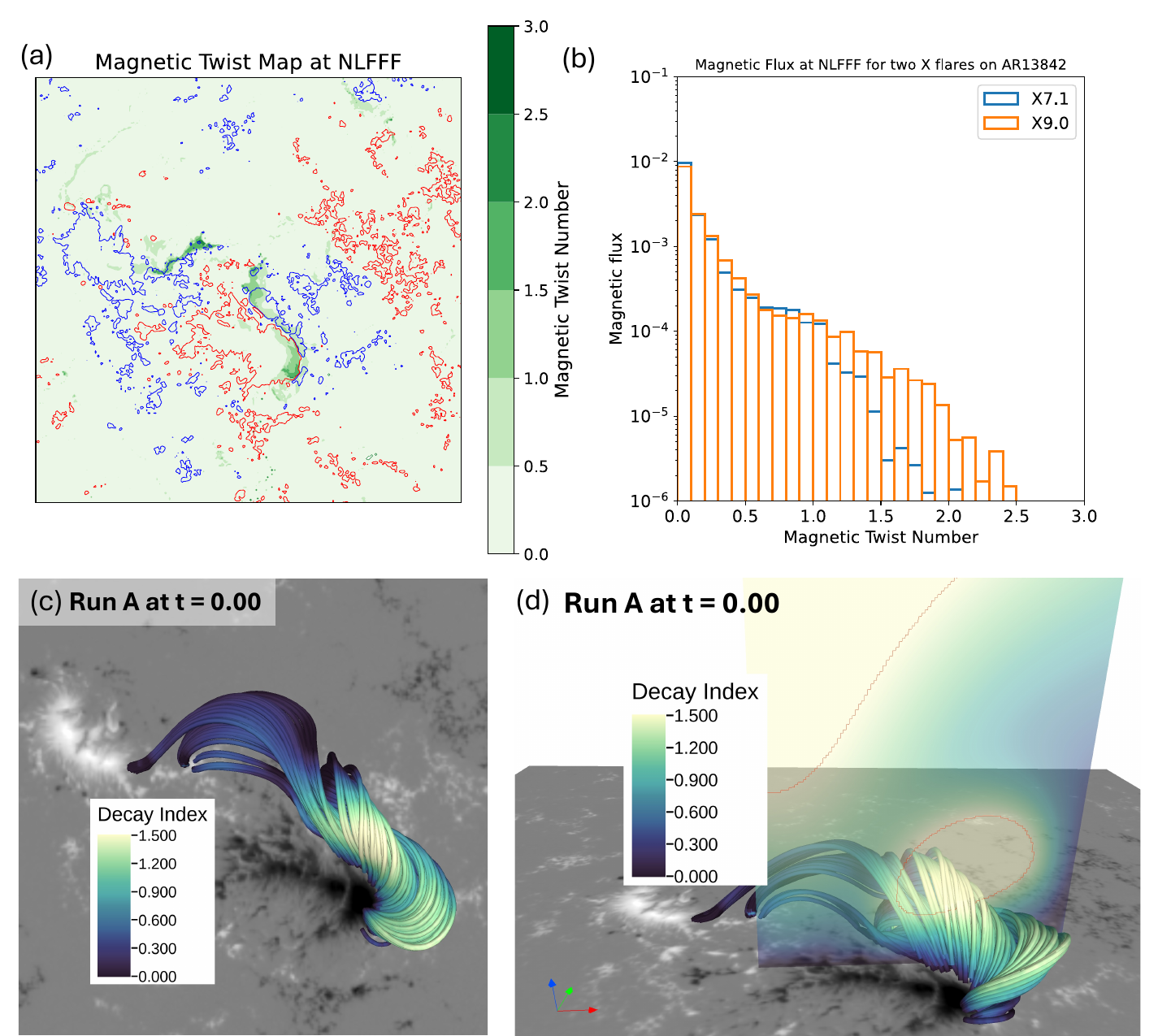}
\caption{(a) Map of the magnetic twist number ($T_w$) at NLFFF. Blue and red contours show the normal component of the magnetic field $B_z=\pm 0.1$ in the simulation, corresponding to $2.87 \times 10^{-2}$ T in the observation. (b) The magnetic flux of NLFFF for the X7.1 flare on October 1, 2024, and the X9.0 flare on October 3, 2024. (c) Twisted field lines with $T_{w}$ exceeding 0.5 in the NLFFF, which correspond to \(t = 0.00\) in Run A. The color of the field lines represents the decay index. (d) Side view of the panel (c) except that the decay index is plotted on the vertical cross-section. The red line shows the decay index $n=1.5$. 
}
\label{fig:fig2}
\end{figure*}

On October 3, 2024, the X9.0 flare was observed in NOAA AR 13842. The data from the Geostationary Operational Environmental Satellite (GOES) indicated that the flare started at 12:08 UT, peaked at 12:18 UT, and ended at 12:27 UT, as shown in Figure \ref{fig:fig1} (a). The CME was observed at 12:48:05 UT by the Solar and Heliospheric Observatory (SOHO)/Large Angle and Spectrometric Coronagraph Experiment (LASCO) C2 telescope (\citealt{Brueckner1995}). Figure \ref{fig:fig1} (b) shows the LASCO/C2 running difference image at 12:53:16 UT. As depicted by the red arrow, the CME was ejected due to the X9.0 flare, where AR 13842 was at the disk center. Finally, this event grew to be observed as a halo CME, accompanied by more ejections from other regions. For more detailed information, please refer to \url{https://cdaw.gsfc.nasa.gov/CME_list/UNIVERSAL_ver2/2024_10/univ2024_10.html}. We used the SDO/HMI vector magnetogram at 10:48 UT (about 1.5 hours before the X9.0 flare) and on the cylindrical equal area (CEA) projection, as the bottom boundary condition to extrapolate the NLFFF. The vector magnetogram in this study was generated in the same way as the hmi.sharp\_cea\_720s (\href{http://jsoc.stanford.edu/ajax/lookdata.html?ds=hmi.sharp_cea_720s}{\texttt{jsoc.stanford.edu}}), except that we expanded the field of view compared to the original sharp\_cea\_720s data to minimize the influence of the size of the simulation box. The size of the HMI map used in this study is about $362.5 \times 362.5~\mathrm{Mm}^2$. Figure \ref{fig:fig1} (c) shows the NLFFF extrapolated from the vector magnetogram of HMI data at 10:48 UT. A sigmoid structure can be observed along the polarity inversion line (PIL) in this panel (c). Figure \ref{fig:fig1} (d) shows the AIA 94 \AA\ image at 10:46:38 UT. Figure \ref{fig:fig1} (e) is superimposed on Figure \ref{fig:fig1} (d) to show the match of the sigmoid structure, emphasizing the reliability of the obtained NLFFF in representing the magnetic configuration leading up to the eruption.

\subsection{Nonlinear Force-free Extrapolation}
\label{sec:sec2.2}
To conduct the NLFFF extrapolation and the data-constrained MHD simulation, we employed the normalized MHD equations from (\citealt{Inoue2014a, Inoue2016}).
\begin{equation}
\mathit{\rho} = \lvert \bm{B} \rvert,
\label{eq:eq1}
\end{equation}
\begin{equation}
\frac{\partial \bm{v}}{\partial t} = -(\bm{v} \cdot \nabla)\bm{v} + \frac{1}{\mathit{\rho}} \bm{J} \times \bm{B} + \mathit{\nu \nabla^2 \bm{v}},
\label{eq:eq2}
\end{equation}
\begin{equation}
\frac{\partial \bm{B}}{\partial t} = \nabla \times (\bm{v} \times \bm{B})+\eta \bm{\nabla}^2\bm{B} - \nabla \bm{\phi},
\label{eq:eq3}
\end{equation}
\begin{equation}
\bm{J} = \nabla \times \bm{B},
\label{eq:eq4}
\end{equation}
\begin{equation}
\frac{\partial \bm{\phi}}{\partial t} + c_h^2 \nabla \cdot \bm{B} = - \frac{c_h^2}{c_p^2} \bm{\phi},
\label{eq:eq5}
\end{equation}
where $\mathit{\rho}$, $\bm{B}$, $\bm{v}$, $\bm{J}$, and $\mathit{\phi}$ are the plasma density, the magnetic flux density, the velocity, the electric current density, and the scalar potential, respectively. The length, the magnetic field, the plasma density, the velocity, the time, and the electric current density are normalized by $L^{*}$ = 362.5 Mm, $B^{*}$ = $2.869 \times 10^{-1}$ T, $\mathit{\rho}^{*}$ (kg m$^{-3}$), where $\mathit{\rho}^{*}$ is the density at the bottom boundary, $V_A^* = {B^*}/{(\mu_0 \rho^*)^{1/2}}$ (m s$^{-1}$), where $\mu_0$ is the magnetic permeability, $\tau_A^* = {L^*}/{V_A^*}$ (s), and $J^* = {B^*}/{\mu_0 L^*}$ (A m$^{-2}$), respectively. We apply the scalar potential $\phi$ in Equation (\ref{eq:eq5}) to reduce the error of ${\bf \nabla }\cdot{\bf B}$ (\citealt{Dedner2002}). The coefficients $\nu$ and $\eta$ are viscosity and electric resistivity adopted for implementing both NLFFF and MHD simulations. We set $\nu=1.0\times 10^{-3}$ and $\eta=5.0\times 10^{-5}+1.0\times 10^{-3}|\bm{J}\times \bm{B}||{\bm v}|^2/|{\bm B}|^2$ in the NLFFF extrapolation. The coefficients ${c_h^2}$ and ${c_p^2}$ in Equation (\ref{eq:eq5}) are fixed at the constant values of 0.04 and 0.1, respectively.

We used the potential field, extrapolated from observed $B_{z}$ using the Green function method (\citealt{Sakurai1982}), as the initial condition. Regarding the boundary conditions, the normal component of the magnetic field is fixed on each boundary surface over time, while the tangential components evolve in accordance with the induction equation, except at the bottom boundary. The velocities are fixed to zero at all boundaries. $\partial / \partial n$ = 0 is applied to $\phi$. Note that the tangential component of the magnetic field at the bottom boundary follows 
\begin{equation}
\bm{B}_{\text{bc}} = \gamma \bm{B}_{\text{obs}} + (1 - \gamma) \bm{B}_{\text{pot}},
\end{equation}
where $\bm{B}_{\text{bc}}$ corresponds to the tangential component, which represents a linear combination of the observed magnetic field ($\bm{B}_{\text{obs}}$) and the potential magnetic field ($\bm{B}_{\text{pot}}$). The parameter $\gamma$ ranges from 0 to 1. Initially, $\gamma$ is set to 0 and is updated as $\gamma + d\gamma$ during each iteration, provided the total Lorentz force $R = \int |\bm{J} \times \bm{B}|^2 dV$ falls below a critical threshold defined as $R_{\text{min}}$. The magnetic field is fixed to the observed value once $\gamma$ reaches 1. We set $R_{{\text{min}}}=5.0 \times 10^{-3}$ and $d\gamma =0.02$. To prevent sharp discontinuities in the velocity, especially between the boundary and inner regions, we impose a velocity limit. Specifically, if the Alfvén Mach number, defined as $v^* \left(= {|\bm{v}|}/{|\bm{v}_{\text{A}}|} \right)$, exceeds a specified limit $v_{\text{max}}$ (which is set to 0.04 in this study), we adjust the velocity using $\bm{v} \rightarrow \left( v_{\text{max}}/v^* \right) \bm{v}$. This method helps to prevent abrupt changes in velocity during the iterative process.

\subsection{Data-constrained MHD Simulation} 
\label{sec:sec2.3}
Subsequently, we conducted data-constrained MHD simulations using the NLFFF as the initial condition as done in \cite{Inoue2014b, Inoue2018}, aiming to capture the evolution from flare initiation to eruption. Although the governing equations are identical to those used in the NLFFF calculation, the primary difference lies in treating the bottom boundary condition for the tangential magnetic field and the removal of the velocity limit. Specifically, the normal component of the magnetic field at the bottom boundary is fixed in time. This treatment corresponds to the assumption that the footpoints of the coronal loops remain stationary, since the photosphere is dominated by plasma pressure and responds on timescales much longer than that of a flare. Therefore, it is reasonable to fix the normal component of the magnetic field in this study. In contrast, the tangential components evolve in the induction equation, with all velocity components set to zero. In these MHD simulations, the resistivity $\eta$ and the viscosity $\nu$ were both fixed at $1.0 \times 10^{-5}$ and $1.0 \times 10^{-4}$, respectively.

For both the NLFFF and MHD simulations, the numerical domain was set to $362.5 \times 362.5 \times 362.5~\mathrm{Mm}^3$, corresponding to $1.0 \times 1.0 \times 1.0$ in dimensionless units. While the original HMI data were sampled at $1000 \times 1000$ grid points, a $2 \times 2$ binning process was applied, resulting in a $500 \times 500$ grid. The simulation time $t$ in this study is normalized by the Alfvén time (see Section \ref{sec:sec2.2}), where $t = 1.00$ corresponds to approximately six minutes in physical time.

  \section{Results}
  \label{sec:sec3}

\subsection{Initial MFR before Eruption}
\label{sec:sec3.1}

\begin{figure*}[ht!]
\plotone{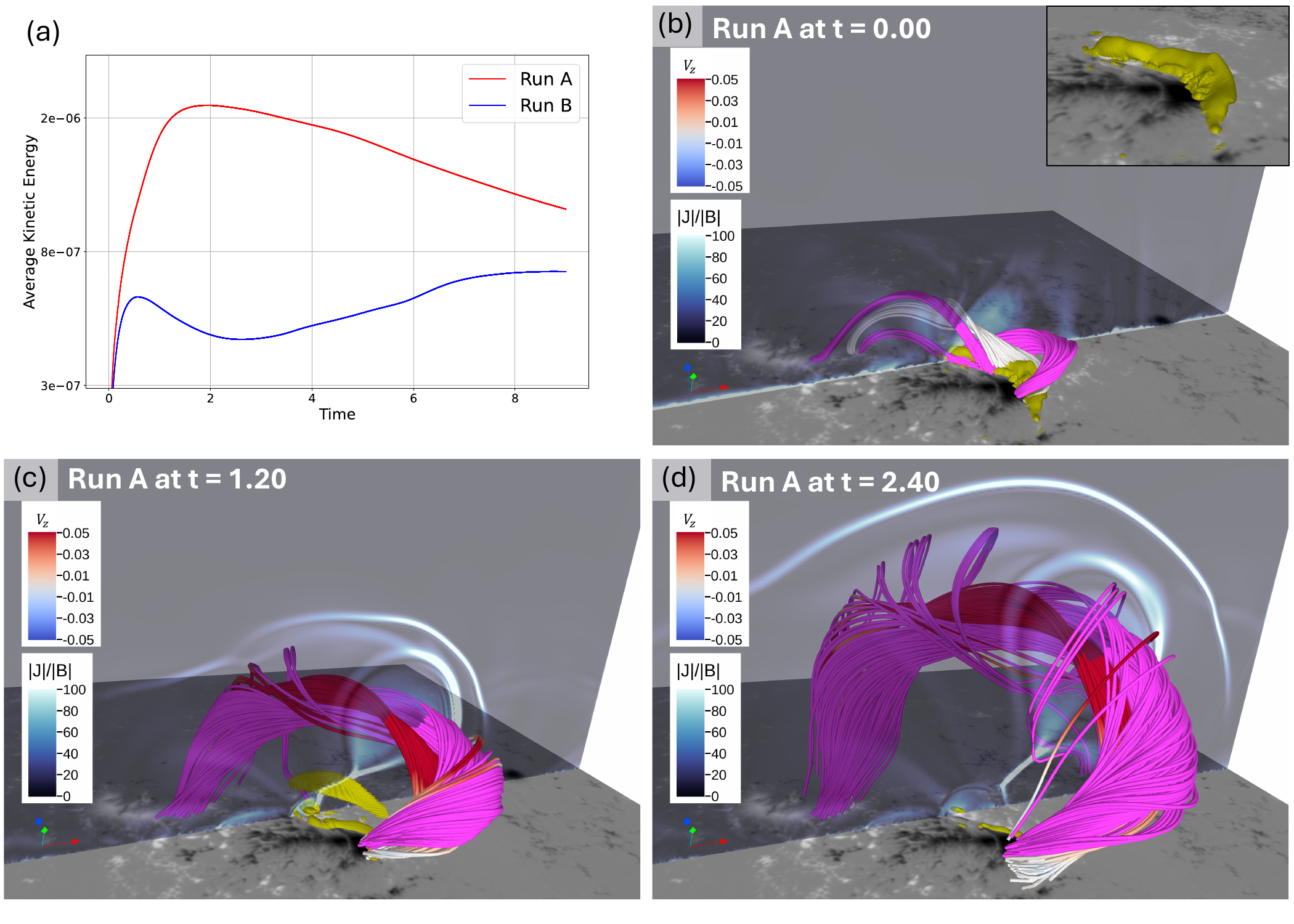}
\caption{(a) Temporal evolution of the averaged kinetic energy for Runs A and B, respectively. (b)--(d) Temporal evolution of the twisted field lines for Run A. The vertical cross-section represents \(|\bm{J}|/|\bm{B}|\). The iso-surface colorer by yellow shows $|\bm{J}|=25$. Purple twisted field lines are part of the sheared field lines in the NLFFF and undergo the tether-cutting reconnection. They ultimately ascend as the MFRs. The field lines colored by \(V_z\) satisfy the decay index of 1.5 or greater at $t=0.00$ and potentially destabilize the TI. The small insertion in the top-right of the panel (b) shows the iso-surface of $|\bm{J}|=25$ at $t=0.00$. An animation of the temporal evolution of Run A is available. The animation proceeds from t = 0.00 to 2.40, and it is provided with two different color maps (left and right). (An animation of this figure is available in the online article.)
}
\label{fig:fig3}
\end{figure*}

\begin{figure*}[ht!]
\plotone{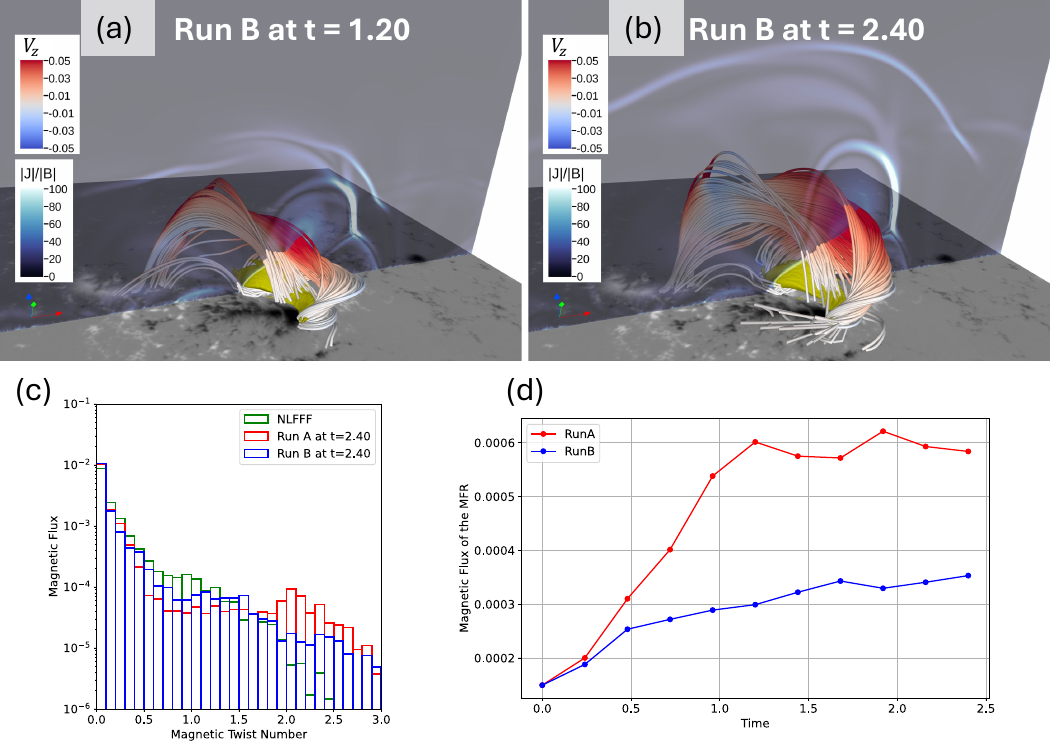}
\caption{(a)--(b) Temporal evolution of the twisted field lines for Run B. The vertical cross-section represents \(|\bm{J}|/|\bm{B}|\). The iso-surface shows $|\bm{J}|=25$. (c) Magnetic flux \(\int |B_z| \, dS\) at NLFFF ($t=0.00$ in Runs A and B) and $t=2.40$ in Runs A and B as a function of $T_w$. (d) Temporal evolution of the magnetic flux of the MFR in Runs A and B, respectively, where the twisted field lines with $T_w > 1.5$ are focused. An animation of Run B (panels (a)-(b)) is available. The animation proceeds from t = 0.00 to 2.40, with Run A on the left and Run B on the right. (An animation of this figure is available in the
online article.)
}
\label{fig:fig4}
\end{figure*}

\begin{figure*}[ht!]
\plotone{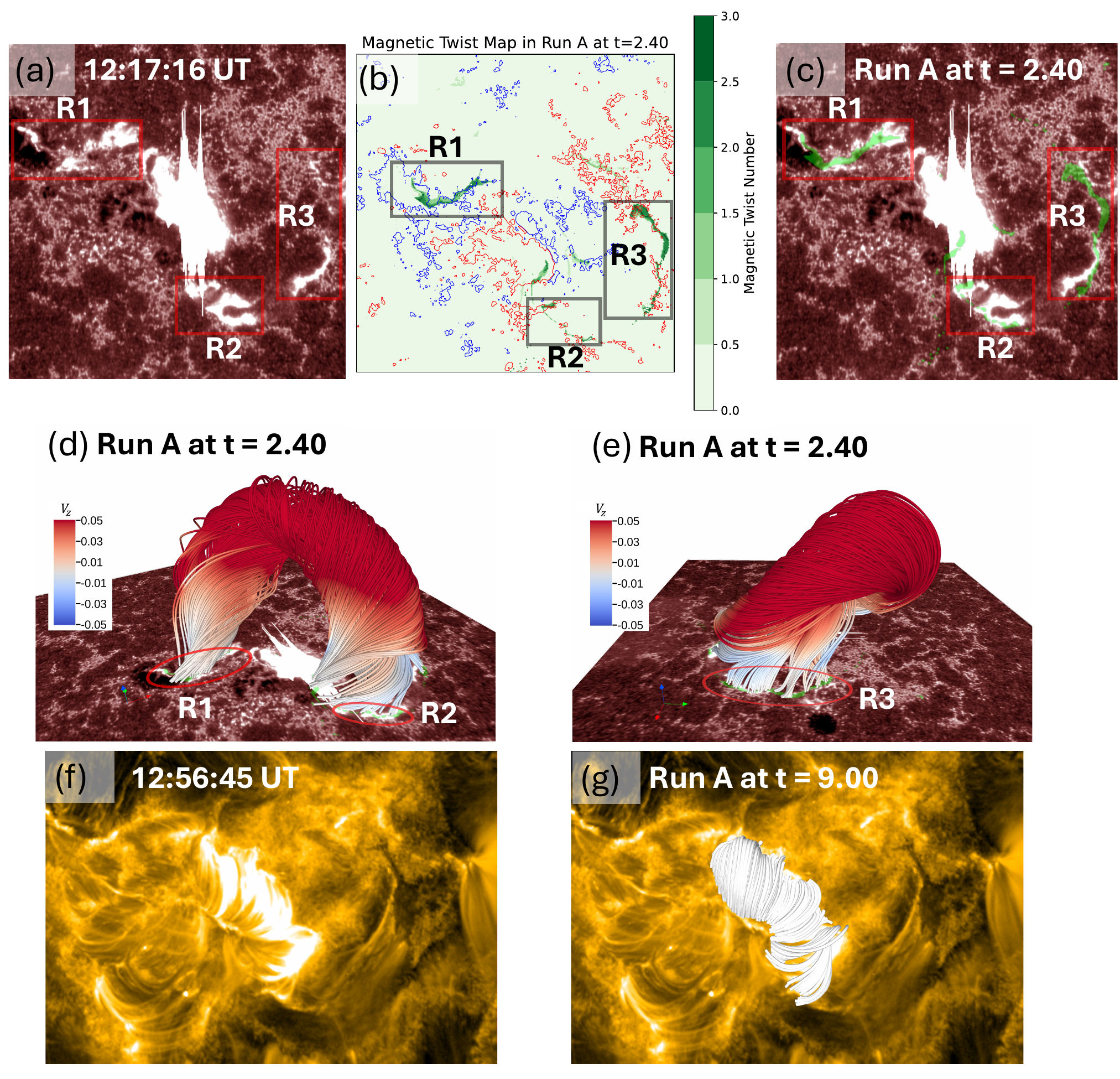}
\caption{ (a) AIA 1700\,\AA\ image at 12:17:16 UT. R1, R2, and R3 show the regions enclosed by the red rectangle as remote brightenings. (b) Map of the magnetic twist number $T_w$ at $t=2.40$ in Run A. Blue and red contours show the normal component of the magnetic field $B_z=\pm 0.1$. (c) Regions where \( T_w \geq 1.5 \) at \( t = 2.40 \) in Run A are filled in green, overlaid on (a). (d) and (e) Twisted field line exceeding $T_w=1.5$ drawn from R1, R2, and R3. The image at the boundary is the same as panel (c). (f) AIA 171\,\AA\ image at 12:56:45 UT. (g) The field lines at \(t = 9.00\) are superimposed on (f).}
\label{fig:fig5}
\end{figure*}

\begin{figure*}[ht!]
\plotone{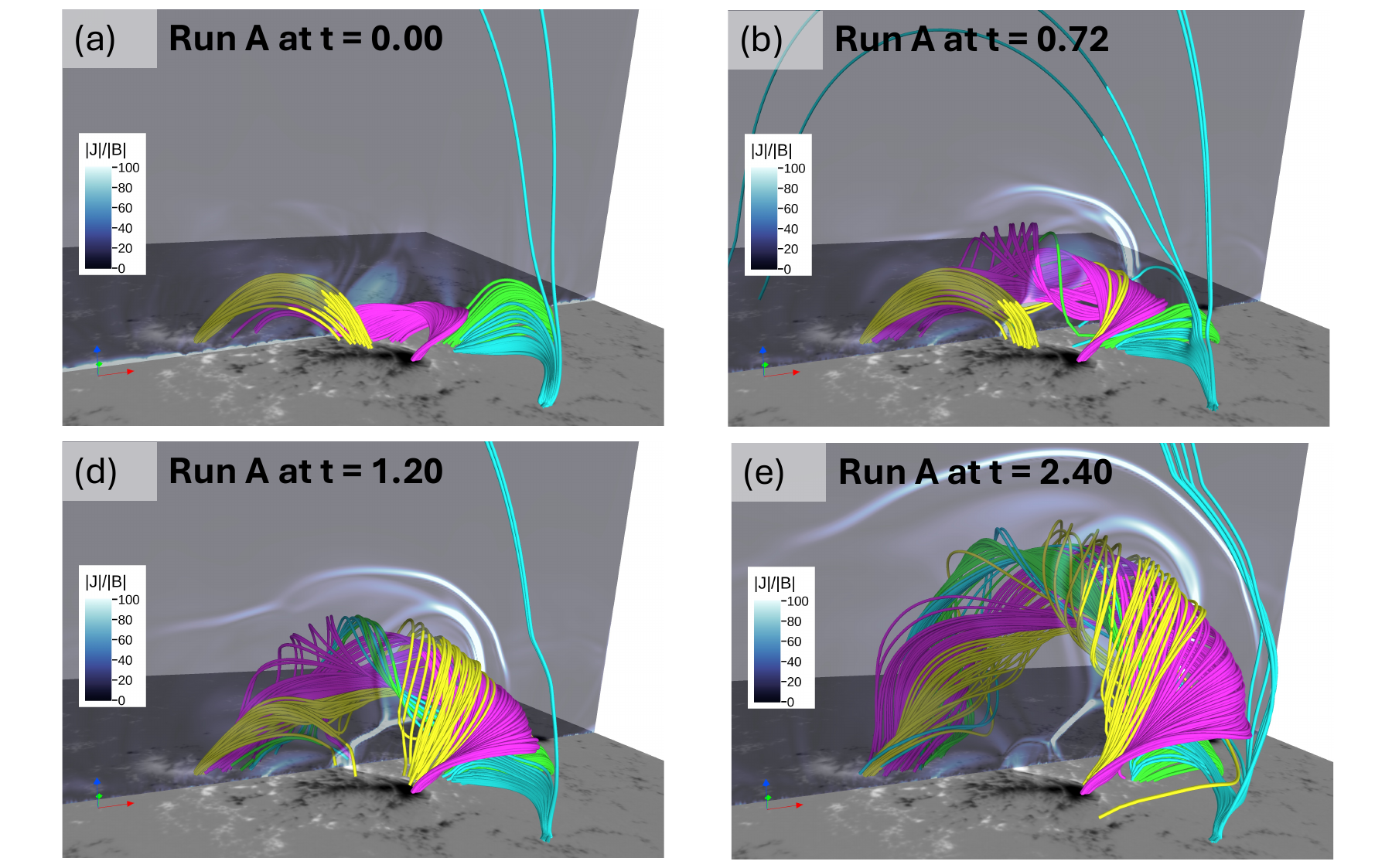}
\caption{(a)-(d) Temporal evolution of the twisted field lines and field lines tracing from R1, R2, and R3 for Run A. Purple field lines are a part of the twisted field lines, shown in Figure \ref{fig:fig3} (b). Yellow, blue, and green field lines are tracing from R1, R2, and R3, respectively. The vertical cross-section represents \(|\bm{J}|/|\bm{B}|\). An animation of the temporal evolution of Run A is available. The animation proceeds from t = 0.00 to 2.40.
\label{fig:fig6}}
\end{figure*}

\begin{figure*}[ht!]
\plotone{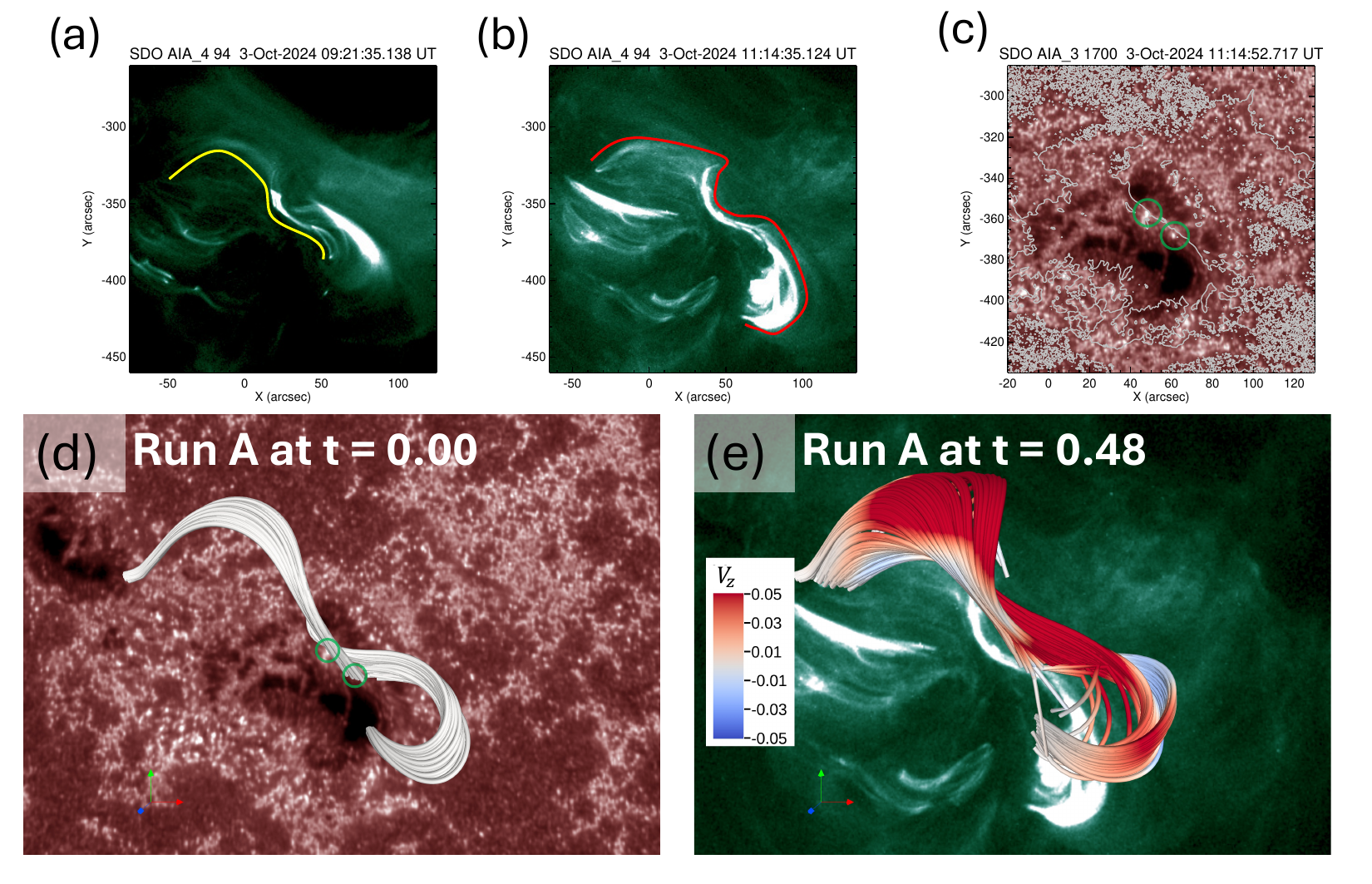}
\caption{ (a) AIA 94\,\AA\ image at 09:21:36 UT. The yellow line indicates a sheared field line, as suggested by the image. (b) AIA 94\,\AA\ image at 11:14:35 UT. The red line indicates the reconnected field lines after the tether-cutting reconnection, as suggested by the image. (c) AIA 1700\,\AA\ image at 11:14:52 UT. Two brightening points enclosed by the green circles suggest the local reconnection. The gray contour indicates the PIL at 11:15:00 UT. (d) Twisted field lines with $T_w > 0.5$ at NLFFF, superimposed on (c). The footpoints of these field lines are located near or overlap with the small brightening points, which are enclosed by green circles and correspond to those shown in panel (c). (e) Field lines at $t=0.48$ in Run A, which were traced from the same points as the magnetic field lines shown in panel (d), superimposed on (b). The field lines are colored by $V_z$.
}
\label{fig:fig7}
\end{figure*}

\begin{figure*}[ht!]
\plotone{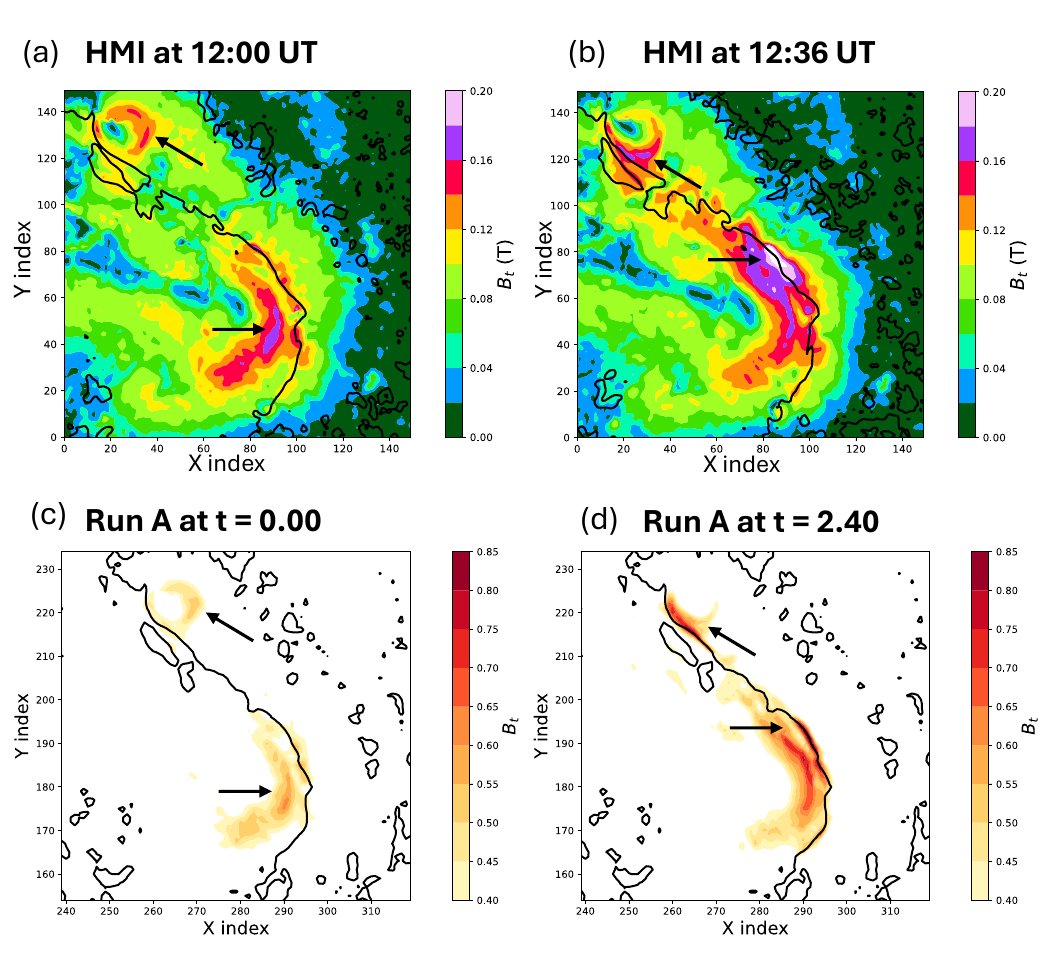}
\caption{Change in the strength of the transverse magnetic fields $B_{t}=\sqrt{B^{2}_x+B^{2}_y}$, which is measured on the photosphere, in the observations and MHD simulations. (a) $B_t$ observed by HMI at 12:00 UT, which is 8 minutes before the starting time of the X9.0 flare. The black lines indicate the PIL. (b) $B_t$ observed by HMI at 12:36 UT, which is 9 minutes after the ending time of the X9.0 flare. (c)-(d) Each $B_t$ is plotted at $t=0.00$ and at $t=2.40$ in Run A, respectively. The black lines indicate the PIL. Black arrows in panels (a)-(d) indicate the enhanced $B_t$ region. An animation of $B_t$ observed by SDO/HMI (panels (a)-(b)) is available. The animated images proceed from 12:00 UT to 12:36 UT. An animation of $B_t$ in Run A (panels (c)-(d)) is available. The animated images proceed from $t=0.00$ to $2.40$. (These animations of this figure are available in the online article.)
\label{fig:fig8}}
\end{figure*}

First, we calculate the magnetic twist $T_w$ in the NLFFF model in the pre-flare state as an indicator of non-potentiality. The magnetic twist is defined as
\begin{equation}T_w = \frac{1}{4\pi} \int \frac{{\bm J}\cdot {\bm B}}{|\bm{B}|^2} \, dl,
\label{eq:eq8}
\end{equation}
where $dl$ is a line element (\citealt{Berger2006}). It is important to emphasize that $T_w$ represents the local twisting between two infinitesimally close field lines, which is different from counting how many turns the field lines perform around the axis of the MFR. Figure \ref{fig:fig2} (a) shows a $T_w$ map with the contours $B_z=\pm 0.1$, corresponding to $\pm 2.87 \times 10^{-2}$ T in the physical unit. The positively twisted field lines run along the PIL, suggesting that the positive helicity is predominant in the active region. Figure \ref{fig:fig2} (b) shows the magnetic flux, \(\int |B_z| \, dS\), for each $T_w$. The X9.0 on October 03, 2024 has more twisted field lines in NLFFF than the X7.1 flare on October 01, 2024. Especially, the amount of the twist that satisfies $T_w \ge 1$ is significantly greater in the former. Therefore, the larger flare was observed 2 days after the X7.1 flare. Furthermore, this result suggests that accumulating the highly twisted field lines with $T_w \ge 1.0$ may be important for causing larger flares.
In Figure \ref{fig:fig2} (c), the field lines in the NLFFF are chosen for those with $T_w$ greater than $0.5$ to trace the dynamics of the eruption. This threshold was motivated by our previous analysis of the X7.1 flare in the same active region (\citealt{Matsumoto2025}).

We calculate the decay index $n$ in NLFFF for TI (\citealt{Bateman1978, Kliem2006}) as
\begin{equation}
n = -\frac{d \ln |\bm{B}_{\text{ex}}|}{d \ln z},
\label{eq:eq9}
\end{equation}
where $\bm{B}_{\text{ex}}$ is the transverse components of the external field, approximated to the potential field in this study. The decay index represents how rapidly the transverse components of the magnetic field decay with height. When the axis of the MFRs exceeds the height of $n=1.5$, the hoop force acting on the MFRs is dominant compared to the suppressing force from the external fields, resulting in an eruption of MFRs. Note that the threshold of $n$ has a range depending on the structure of the MFR and the dynamics in the boundary states (\citealt{Olmedo2010, Ju2018, Zuccarello2015}). Figure \ref{fig:fig2} (c) shows the twisted field lines with $T_w > 0.5$ have the potential to become unstable to TI, as a part of the MFR exceeds the height at which the decay index reaches 1.5. However, a potentially stable region with $n\approx 1.0$ is present above the twisted field lines, as shown in Figure \ref{fig:fig2} (d). Therefore, to fully understand the evolution of the magnetic field, we conduct an MHD simulation.

\subsection{Modeling with Full Reconnection Onset}
\label{sec:sec3.2}
To reveal the dynamics of 3D magnetic field lines, we performed a data-constrained MHD simulation using the NLFFF as the initial condition. We refer to this MHD simulation as Run A. Figure \ref{fig:fig3} (a) shows the temporal evolution of the kinetic energy ($\int (\rho |v|^2 / 2) \, dV$, where $dV$ is a volume element) for Runs A and B. We will discuss Run B later. Figures \ref{fig:fig3} (b)-(d) show the temporal evolution of the twisted field lines in Run A, which are selected from Figure \ref{fig:fig2} (c) for easier viewing. The tether-cutting reconnection occurred in the purple field lines in Figures \ref{fig:fig3} (b) and (c). The colored twisted field lines with $V_z$ are depicted above the purple field lines. Note that the induction equation (Equation (\ref{eq:eq3})) includes electric resistivity, which induces magnetic reconnection in the region of strong current. Section \ref{sec:sec3.1} already suggested that these field lines are either unstable or nearly unstable to the TI. Either way, Figure \ref{fig:fig3} (d) shows that the newly formed MFR (colored in purple) generated by tether-cutting reconnection pushes these field lines upward, accelerating their destabilization. The purple field lines added a twist to the erupting MFR and reinforced its structure.

\subsection{Modeling with Suppressed Reconnection}
\label{sec:sec3.3}
In section \ref{sec:sec3.2}, we found that tether-cutting reconnection and TI are essential in triggering the X9.0 flare. However, the precise role of these processes during the eruption remains unclear. To investigate this point further, we conducted an additional simulation, referred to as Run B, in which magnetic reconnection was artificially suppressed by setting the velocity to zero in regions with strong current density (\citealt{Inoue2006, Inoue2018, Yamasaki2022, Matsumoto2025}). The suppression of reconnection was applied specifically to areas with \(|\bm{J}| > 25\), which is visualized as an iso-surface in the small inset of Figure \ref{fig:fig3} (b). It should be noted that this suppression was localized and did not completely stop the reconnection in Run B.

Figures \ref{fig:fig4} (a) and (b) show the temporal evolution of the magnetic field lines in Run B. The selected field lines are traced from the same location as in Run A but colored by $V_z$. The tether-cutting reconnection was clearly suppressed compared to Figures \ref{fig:fig3} (c) and (d). Figure \ref{fig:fig4} (c) shows the magnetic flux at NLFFF ($t=0.00$) and $t=2.40$ in Runs A and B as a function of $T_w$. Throughout the reconnection, the magnetic flux with $T_w \gtrsim  1.5$ increased by $t=2.40$, while the magnetic flux occupied with $0.5 < T_w < 1.5$ decreased due to the conversion to erupting field lines with the twist $T_w \gtrsim  1.5$.
From this result, since the conversion occurs more prominently in Run A than in Run B, it clearly indicates that suppressing the reconnection prevents the formation of a stronger MFR. 
To evaluate the magnetic flux of the highly twisted field lines, we calculated the magnetic flux \(\Phi = \int_{T_w > 1.5} |B_z| \, dS\), where \(B_z\) corresponds to the normal component at the bottom boundary. The suppression of MFR growth in Run B is clearly visible in Figure \ref{fig:fig4} (d). As shown in Figure \ref{fig:fig3} (a), the reconnection occurred after $t=3.0$ in Run B, and the kinetic energy increased.  One possible reason is that the reconnection, which was not suppressed in the region with \(|\bm{J}| < 25\),  may have contributed to the acceleration of the MFR. Nevertheless, the kinetic energy in Run B is much smaller than Run A. Therefore, the tether-cutting reconnection in the pre-eruption (by $t=1.20$ in this study) is essential in the rapid eruption. When reconnection occurs between sheared field lines, twisted field lines are formed, resulting in an enhancement of the toroidal current in the MFR and strengthening the hoop force. Consequently, the tether-cutting reconnection in the pre-eruption is a key factor in driving the subsequent eruption through torus instability as discussed in \citealt{Matsumoto2025}.

\subsection{Comparison with the Observations}
\subsubsection{Remote Brightenings at 1700 \AA\ and Flare Loop}
To clarify the dynamics seen in the observation, we compared the simulation results with observations. Figure \ref{fig:fig5} (a) shows the AIA 1700 \AA\ image at 12:17:16 UT. In addition to the brightening at the center of the image, corresponding to the flare ribbons, three remote brightening regions were observed as R1, R2, and R3. Figure \ref{fig:fig5} (b) shows a $T_w$ map at $t=2.40$ in Run A. There correspond to the locations at which the twisted field lines are anchored. In Figure \ref{fig:fig5} (c), the green regions exceeding $T_w=1.5$ in Figure \ref{fig:fig5} (b) are superimposed onto Figure \ref{fig:fig5} (a), showing that the remote brightening regions coincided with the footpoints of the twisted field lines. Figures \ref{fig:fig5} (d) and (e) show the erupting field lines exceeding twist number $T_w=1.5$, whose footpoints are anchored in regions R1, R2, and R3. Therefore, this result indicates that the remote brightening regions observed at 1700 \AA\ correspond to the footpoint locations of erupting MFR. Figure \ref{fig:fig5} (f) shows the AIA 171 \AA\ image at 12:56:45 UT. The brightening in the center of Figure \ref{fig:fig5} (c) corresponds to the flare loops in Figure \ref{fig:fig5} (f). In this study, the MHD simulation reproduced the flare loops in Figure \ref{fig:fig5} (g). 
\subsubsection{Detailed dynamics associated with the remote brightenings R1, R2, and R3}

In NLFFF, the footpoints of the twisted field lines didn't exist on R2 and R3, as shown in Figure \ref{fig:fig2} (a). However, Figure \ref{fig:fig5} (b) shows that the footpoints of the twisted field lines exist at the R2 and R3 in the MHD simulation. Figure \ref{fig:fig6} shows the temporal evolution of the twisted field lines and field lines anchored in R1, R2, and R3 for Run A. Figures \ref{fig:fig5} (b) and \ref{fig:fig6} show that the pre-existing twisted region at R1 expanded eastward, suggesting an eastward drift of its footpoints due to reconnection with adjacent loops (\citealt{Aulanier2019}). Similarly, the twisted region at R2 and R3 appeared as the expanding MFR reconnected with and incorporated the overlying coronal loops anchored there (\citealt{Gibson2008}). This interaction not only established new twisted footpoints in these regions but also assisted in strengthening the overall twist of the erupting MFR, complementing the primary role of tether-cutting reconnection in producing the highly twisted flux rope, as discussed in Section \ref{sec:sec3.3}.
\subsubsection{Tether-cutting Reconnection before the X9.0 Flare}
\label{sec:sec3.4.2}
Figure \ref{fig:fig7} (a) shows the AIA 94 \AA\ image at 09:21:35 UT, which is about 2.5 hours before the flare. In the observation, sheared field lines were seen. To make this clearer, we overlaid a yellow line for better visualization.  Figure \ref{fig:fig7} (b) shows the AIA 94 \AA\ image at 11:14:35 UT, which is about 1 hour before the flare. In Figure \ref{fig:fig7} (b), we have added red lines to help visualize the newly formed field lines. These lines suggest that the sheared field lines seen in Figure \ref{fig:fig7} (a) underwent tether-cutting reconnection, resulting in the formation of new field lines. 

The AIA 1700 \AA\ image at 11:14:52 UT reveals two brightening points along the PIL, indicated with the green circles in Figure \ref{fig:fig7} (c). These brightening points persisting from 11:08:05 UT to 11:23:53 UT suggest the occurrence of localized reconnection at those positions. Figure \ref{fig:fig7} (d) shows the twisted field lines with $T_w >0.5$, whose footpoints are found either within or very close to the two brightening regions. To show the evolution of the field lines, we plot the field lines at $t=0.48$ in Figure \ref{fig:fig7} (e), for comparison with the initial field lines shown in Figure \ref{fig:fig7} (d). Namely, the sheared field lines in Figure \ref{fig:fig7} (d) underwent tether-cutting reconnection to form the twisted field lines in Figure \ref{fig:fig7} (e), in good agreement with the observation in Figure \ref{fig:fig7} (b). Therefore, these results demonstrate that our simulation is capable of capturing the early evolution of the twisted field lines. The reconnected loops as shown in Figure \ref{fig:fig7} (b) were observed around 09:54 UT, prior to the appearance of the two brightening points in Figure \ref{fig:fig7} (c). This suggests that the tether-cutting reconnection had already started at 09:54 UT, and the reconnection associated with the two brightening points around 11:00 UT contributed to the further development of the MFR.

\subsubsection{Enhancement of Transverse Magnetic Field after the X9.0 Flare}

Figures \ref{fig:fig8} (a) and (b) show the strength of the transverse magnetic fields, $B_{t}=\sqrt{B^{2}_x+B^{2}_y}$, observed by SDO/HMI at 12:00 UT and 12:36 UT. The black lines indicate the PIL. In Figure \ref{fig:fig8} (a), the regions of strong $B_t$ are distributed slightly off the PIL, appearing as two separate concentrations to the north and south. As shown by the black arrows in Figures \ref{fig:fig8} (a) and (b), the strong $B_t$ region in the northern part has shifted onto the PIL, while the region in the southern part has expanded. In both regions, the field strength has been further enhanced by 12:36 UT. Figures \ref{fig:fig8} (c) and (d) show $B_t$ at the bottom boundary at $t=0.00$ and $t=2.40$ in Run A. The black lines indicate the PIL at the bottom boundary. The MHD simulation successfully reproduced the changes in the location and strength of $B_t$ after the flare, showing good agreement with the observations. The accumulation of the loops by tether-cutting reconnection enhances the transverse field (\citealt{Hudson2008, Wang2010, Wang2012, Liu2012, Nian2023}). Since the evolution of the transverse components follows the induction equation at the bottom boundary, it no longer has to follow the observation. Nevertheless, the data-constrained MHD simulation successfully reproduced the enhancement of the transverse components as observed, as \citealt{Bian2023} demonstrated a similar enhancement in their MHD simulation as part of a modeling approach.

\section{Discussion}
\label{sec:sec4}
\begin{figure*}[ht!]
\plotone{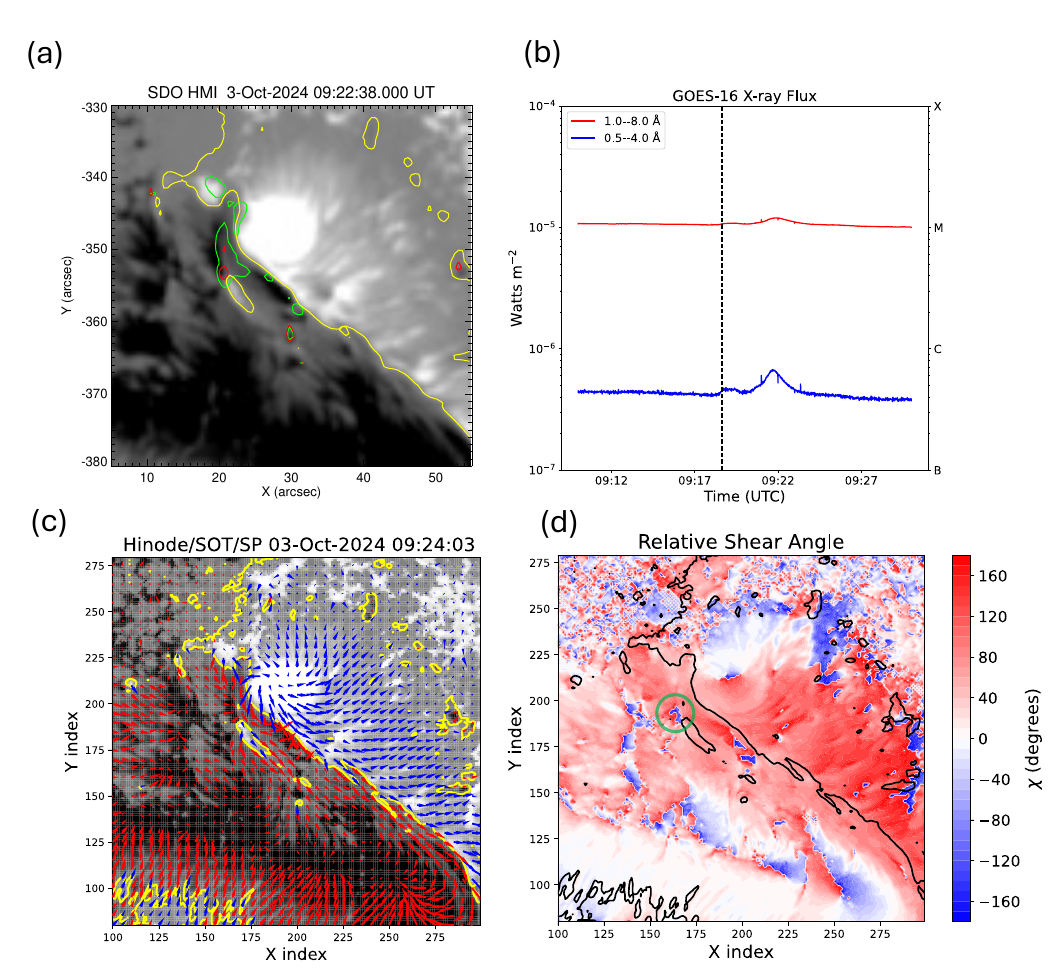}
\caption{(a) Magnetogram observed by SDO/HMI at 09:22:38 UT. The yellow lines indicate the PIL. The green contours indicate the intensity of 300 DN of AIA 1600 \AA\ at 09:19:26 UT. The red contours indicate the intensity of 3000 DN of AIA 1700 \AA\ at 09:19:40 UT. (b) Time evolution of the X-ray flux measured by the GOES-16 satellite between 09:10 UT  and 09:30 UT on October 3. The black vertical dashed line indicates 09:18:38 UT when the brightening of AIA 1600 \AA\ increased the intensity. (c) Magnetogram observed by Hinode/SOT/SP at 09:24:03. The yellow lines indicate the PIL. Red and blue arrows indicate the transverse magnetic fields at the negative and positive $B_z$, respectively. (d) Map of the relative shear angle $\chi$, showing the sign of the magnetic helicity. The green circle indicates one of the regions dominated by the negative helicity, corresponding to the AIA brightenings in panel (a).
\label{fig:fig9}}
\end{figure*}

\begin{figure*}[ht!]
\plotone{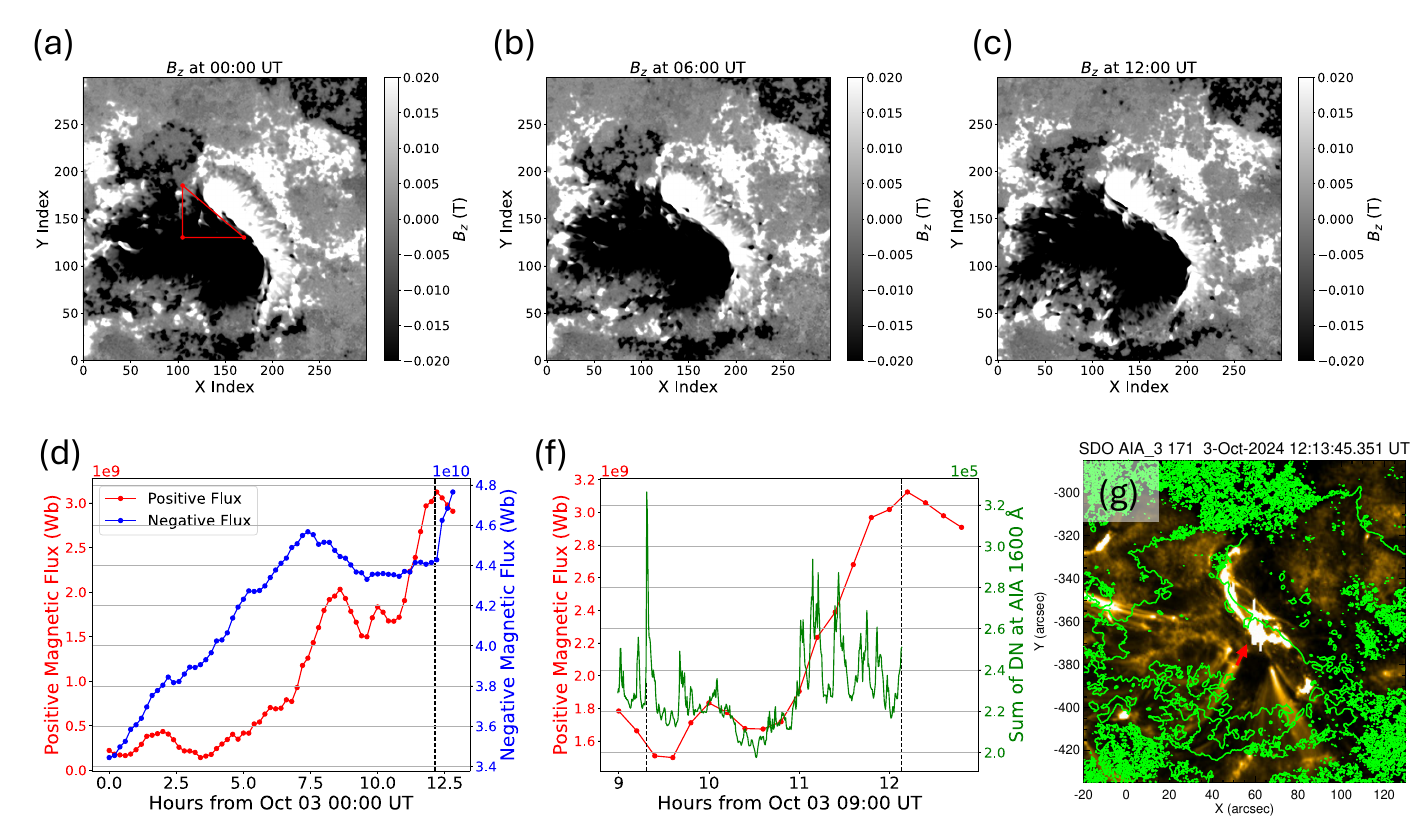}
\caption{(a)-(c) Temporal evolution of $B_z$ observed by SDO/HMI from 00:00 UT to 12:00 UT on October 03. The red triangle in panel (a) shows the region to calculate the temporal evolution of positive/negative magnetic flux in panel (d). (d) Temporal evolution of the positive and negative magnetic flux observed by SDO/HMI between 00:00 UT on 2024 October 3 and 12:48 UT on October 3, respectively. The black dotted vertical line indicates the time 12:08 UT, which is the starting time of the X9.0 flare. (f) Temporal evolution of the positive magnetic flux as shown in panel (d) and total DN at 1600 \AA\ between 08:59:26 UT and 12:08:14 UT, respectively. The black dotted vertical lines indicate 09:18:38 UT, suggesting the local reconnection, and 12:08 UT, respectively. (g) AIA 171 \AA\ image at 12:13:45 UT. The green lines represent the PIL at 12:13:30 UT. The red arrow indicates the brightening point, suggesting the start of the X9.0 flare. An animation of SDO/HMI (panels (a)-(c)) is available. The animated images proceed from 00:00 UT to 12:48 UT. (An animation of this figure is available in the online article.)
\label{fig:fig10}}
\end{figure*}

\subsection{Possible Triggering Process suggested by Hinode/SOT/SP}
\label{sec:sec4.1}
In our MHD simulations, the reconnection automatically started at the strong current region due to the resistivity in the induction equation. Therefore, we can not assume that this reconnection is the triggering process from the simulation results alone. In this section, we attempt to reveal the triggering process by using the Hinode/SOT/Spectro-Polarimeter (SP) to detect the fine-scale magnetic fields in combination with SDO/HMI. The data of Hinode/SOT/SP is obtained by \textit{sotsp\_getdata.pro}  in the
Solar Soft-Ware (SSW) package. The $180^{\circ}$ ambiguity in the transverse fields is solved using ME0 code (\citealt{Leka2009}). In Figure \ref{fig:fig9} (a), a small positive magnetic field region can be seen at ($x, y$) = (23 arcsec, -357 arcsec), slightly offset from the main PIL. The green contours indicate the 300 DN (data number) intensity level of 1600 \AA\ at 09:19:26 UT, while the red contours represent the 3000 DN intensity level of 1700 \AA\ at 09:19:40 UT. These bright regions, seen in both AIA 1600 and 1700 \AA\ images, are located just north of the small positive magnetic field region. Notably, the intensity of 1600 \AA\ started to increase after 09:18:38 UT, indicating the onset of local brightening at that time. As shown in Figure \ref{fig:fig9} (b), the soft-X ray increased after this time, suggesting that the reconnection might happen around this time. The bright region at 1600 \AA\ suggests that tether-cutting reconnection occurred between pre-existing field lines. Since the brightening extends not only around the small positive magnetic field region but also further to the north, the reconnection likely took place slightly above the photosphere rather than being confined to the local region near the bottom. Another possibility is that reconnection related to flux cancellation may be involved, as flux cancellation has been reported at the same location during this time period (\citealt{Ding2025}). Figure \ref{fig:fig9} (c) shows the magnetogram observed by Hinode/SOT/SP at 09:24:03 UT. The yellow lines indicate the PIL, while the red and blue arrows represent the transverse magnetic fields in regions of negative and positive $B_z$, respectively. Along the main PIL, the magnetic field lines are strongly sheared. To discuss the type of triggering process associated with the small magnetic structure, suggested by \citealt{Kusano2012}, we calculated the relative shear angle, defined as $\chi$,  between the potential field and the observation by the following equation. 
\begin{equation}
 \chi = \tan^{-1} \frac{ |\bm{B}_p \times \bm{B}_t|}{\bm{B}_p \cdot \bm{B}_t},
\end{equation}
where $ \bm{B}_{\text{p}}$ is the transverse component of the potential field and $ \bm{B}_{\text{t}}$ is that of the observational data. The positive (negative) $\chi$ corresponds to the positive (negative) helicity. Figure \ref{fig:fig9} (d) shows the distribution of $\chi$ using the Hinode/SOT/SP data. Around the main PIL, the positive helicity is dominant, which is consistent with the result shown in Figure \ref{fig:fig2} (a). However, the negative helicity can be seen within the region, for instance, enclosed by the green circle, located to the north of the small positive magnetic field region. This region has $\chi \approx - 90^\circ$, while surrounding regions have $\chi \approx 90^\circ$. This suggests the reverse-shear (RS) type described in \citealt{Kusano2012}, in which the orientation of the small magnetic field region is nearly opposite to that of the major polarity. Furthermore, this location with the negative $\chi$ corresponds to the brightening point at 1700 \AA\ shown in Figure \ref{fig:fig9} (a). Therefore, the RS-type had the potential to trigger the X9.0 flare and might enhance the local reconnection at the bottom. Note that the small magnetic structure is on the main PIL in \citealt{Kusano2012}, which is inconsistent with the observation in Figure \ref{fig:fig9} (a). However, the RS-type magnetic field could trigger a flare even though it was located slightly away from the PIL (\citealt{Bamba2017}). Although the RS-type structure was observed approximately 2.5 hours before the X9.0 flare, it did not immediately lead to the onset of the flare. This suggests that the RS-type structure's role was likely to trigger the tether-cutting reconnection rather than acting as the sole trigger of the X9.0 flare, which aligns with the finding that other conditions are required to produce a major flare (\citealt{Bamba2013}).

\subsection{Possible Triggering Process suggested by SDO/HMI}
\label{sec:sec4.2}
Flux emergence is one of the candidates for triggering the flare. Regarding flux emergence, we investigated the evolution of the small magnetic field, which was discussed in the previous section. Figures \ref{fig:fig10} (a)-(c) show the evolution of $B_z$ observed by the SDO/HMI from 00:00 UT to 12:00 UT on October 3. Inside the red triangle in Figure \ref{fig:fig10} (a), the flux emergence was observed in time. Using hmi.sharp\_cea\_720s data, we calculated the magnetic flux for the positive and negative magnetic fields inside the red triangle as shown in Figure \ref{fig:fig10} (a). Figure \ref{fig:fig10} (d) shows the temporal evolution of the magnetic flux from 00:00 UT to 12:48 UT. The positive and negative polarities have steadily increased since 00:00 UT, indicating ongoing flux emergence. The positive polarity suddenly increases about 2 hours before the flare and peaks just before the X9.0 flare and then declines, suggesting that sufficient flux emergence was crucial for triggering the flare. 

Next, to investigate the relationship between the flux emergence and the reconnection, we traced the evolution at 1600 \AA\ . Specifically, since we needed to track the active region, we performed the correlation tracking. We focused on the region spanning from $x = 11.6$~arcsec to $36.0$~arcsec and from $y = -370.2$~arcsec to $-333.7$~arcsec at 08:59:26~UT, which includes the region where we discussed the reconnection in Section~\ref{sec:sec4.1} as a subset. Figure \ref{fig:fig10} (f) shows the temporal evolution of total DN at 1600 \AA\ and the positive magnetic flux, which is the same as Figure \ref{fig:fig10} (d). After 10:30 UT, the positive magnetic flux increased continuously, during which brightening at 1600 \AA\ increased many times until the starting time of the X9.0 flare, as indicated by the vertical line in Figure \ref{fig:fig10} (f). This suggests that the continuous flux emergence disturbs the pre-existing magnetic fields in the solar corona, resulting in the reconnection with them as observed at 1600 \AA\ . Regarding the brightening at 09:18:38 UT in Figure \ref{fig:fig10} (f), this is not associated with the flux emergence, as mentioned in section \ref{sec:sec4.1}. Figure \ref{fig:fig10} (g) shows the AIA 171 \AA\ image at 12:13:45 UT. The location of the start of the X9.0 flare is not on the PIL, but it has slightly shifted to the region of flux emergence. This suggests that the flux emergence should be a critical factor in triggering the X9.0 flare, although it may not be the sole driver. The emergence, which persisted from around 00:00 UT until the flare onset, was accompanied by local reconnection involving the RS-type structure (Section \ref{sec:sec4.1}) and, later, tether-cutting reconnection between sheared field lines (Section \ref{sec:sec3.4.2}). These successive reconnection processes contributed to the pre-eruptive magnetic configuration. Ultimately, the flare was triggered by magnetic reconnection between the emerging flux and the ambient fields.

\subsection{Flux Emergence: Trigger or Buildup?}
\label{sec:sec4.3}
The role of the sustained flux emergence observed before the X9.0 flare requires careful interpretation, as it is possible that it served as either a direct trigger or part of a long-term energy buildup process. The trigger hypothesis is supported by the spatial and temporal correlation between the emergence and the flare onset, as discussed in Section \ref{sec:sec4.2}. The buildup hypothesis, on the other hand, is supported by the sustained 12-hour duration of the emergence, suggesting that the emergence is needed to form the pre-eruptive magnetic structure as the injection of energy (\citealt{ Nitta1996,Sun2012}). Distinguishing between these two roles is challenging. Since our data-constrained MHD simulation cannot model flux emergence, it may still point to the possibility that it played a role as a triggering mechanism.

\section{Summary}
\label{sec:sec5}

Using data-constrained MHD simulations, we investigated the 3D dynamics of the X9.0-class solar flare on 2024~October~3 in NOAA AR~13842 from the initiation to the eruption. Our simulation reproduced the dynamics of the magnetic field associated with X9.0 flare, including the tether-cutting reconnection, the formation of the MFR, and the observed flare morphology. We summarize the findings below. 
\begin{itemize}
\item We found that the tether-cutting reconnection in the pre-eruption phase is crucial in accelerating the MFRs. The tether-cutting reconnection initiates the flare, generating twisted field lines, strengthening the toroidal current inside the MFRs, and amplifying the upward hoop force. Thus, how electric current builds up in the MFRs in the pre-eruption phase is key to understanding the onset of the major eruptions. These dynamics are similar to those seen in X7.1 flare reproduced by the MHD simulation in \citealt{Matsumoto2025}.

\item In addition to the typical two flare ribbons, the X9.0 flare produced remote brightening, which is distant from the main PIL. Our simulation reveals that the MFR and the reconnected flare loops are anchored at different locations. The remote brightening observed at 1700 \AA\ corresponds to the footpoints of the MFR, whereas the flare loops are rooted in the flare ribbons. This topological distinction is explicitly identified in our model. Furthermore, our simulation shows these remote footpoints are established as the erupting MFR reconnects with and incorporates surrounding coronal loops, strengthening the MFR's twist.

\item Time sequence of AIA 94 \AA\ images shows the tether-cutting reconnection, and the small brightening at 1700 \AA\ suggested the reconnection occurred there. Our simulation reproduced this tether-cutting reconnection and showed that the footpoints of reconnected field lines are on these brightening regions as observed at 94 \AA\ and 1700 \AA\ . 

\item The enhancement of the transverse magnetic fields observed along the PIL after the X9.0 flare could also be reproduced by our data-constrained MHD simulation. This can be due to the contraction of the magnetic fields by the tether-cutting reconnection and the downward Lorentz force acting on the photosphere.

\item Using Hinode/SOT/SP data, we found the flare (RS-type) around the small positive magnetic region. The simultaneous increase of EUV flux at 1600 \AA\ suggests the tether-cutting reconnection. The location of the brightening region at 1700 \AA\ is the same as that of the RS-type, suggesting that the reconnection associated with the RS-type occurred there. Note that this structure was observed about 2.5 hours before the X9.0 flare. It suggests its role was likely to trigger the tether-cutting reconnection rather than acting as the sole trigger for the eruption.

\item We observed the emergence of the small positive magnetic flux that continued for 12 hours until the X9.0 flare. We also observed an increase of brightening was observed at 1600~\AA\, coinciding with the increase in magnetic flux. This correlation suggests that the emerging flux interacts with the coronal magnetic field and may have contributed to the magnetic reconnection. Furthermore, the brightening in the AIA 171~{\AA} image at the onset of the flare was not located along the PIL, but relatively close to the emerging flux region. This spatial correspondence further suggests the scenario in which the emerging flux acted as the trigger for the X9.0 flare.

\end{itemize}
These dynamic processes leading to the eruption associated with the X9.0 flare could be reproduced by the data-constrained MHD simulation. However, this is primarily due to the resistivity working in a strong current sheet, which triggered magnetic reconnection. As a result, the initial equilibrium state represented by the NLFFF could not be maintained, ultimately leading to the eruption. Therefore, what triggers the reconnection becomes a crucial question. A key similarity between the present result and the findings of the X7.1 flare in \citealt{Matsumoto2025} lies in the eruption dynamics, while the primary difference is the degree of accumulated magnetic twist. Although the flux emergence and the associated successive reconnection processes are critical in triggering the X9.0 flare, they were inferred from observational data and are not explicitly modeled in our simulation. Therefore, whether the emergence acted as a direct trigger or as part of the energy buildup process remains an open question. In future work, it will be necessary to verify whether the triggering process proposed in this study can indeed lead to solar flares.

\begin{acknowledgments}
We are grateful to the anonymous reviewer for the constructive comments, which helped to improve the paper. We also thank Prof. Kanya Kusano for the useful discussion. This study is supported by NASA grants 80NSSC23K0406, 80NSSC21K1671, 80NSSC21K0003, 80NSCC24M0174, and NSF grants AST-2204384,  AGS-2145253, 2149748, 2206424, 2309939 and 2401229
  The 3D visualizations were produced using VAPOR (\href{http://www.vapor.ucar.edu}{\texttt{www.vapor.ucar.edu}}), a product of the National Center for Atmospheric Research (\citealt{Li2019}). All calculations of the MHD simulations in this study were performed using the computing facilities of the High Performance Computing Center (HPCC) at the New Jersey Institute of Technology. A part of this study was carried out using the computational resources of the Center for Integrated Data Science, Institute for Space-Earth Environmental Research, Nagoya University.
\end{acknowledgments}

\bibliography{sample631}{}
\bibliographystyle{aasjournal}
%




\end{document}